\documentclass[11pt,aps,prd,amsmath,amssymb,superscriptaddress,nofootinbib]{revtex4-1}
\pdfoutput=1
\usepackage{amsmath,amssymb} 
\usepackage{epsfig} 
\usepackage{color}
\usepackage{bm} 
\usepackage{rotating}
\usepackage{physics}
\usepackage{braket}
\usepackage{multirow}
\usepackage{hhline}
\usepackage{colortbl}
\usepackage[table]{xcolor}
\usepackage{booktabs}
\usepackage{graphicx}
\usepackage{slashed} 
\usepackage[left=2.5cm,right=2.5cm,top=2cm,bottom=2cm]{geometry}

\newcommand{\X}{X(3872)}
\newcommand{\mbare}{\stackrel{\circ}{m}_{c\bar c}}

\begin{document}

\title{The effects of charmonium on the properties of the $1^{++}$ hidden charm poles in effective field theory}
\author{E. Cincioglu}
\email{elif.cincioglu@gmail.com}
\affiliation{Department of Physics, Middle East Technical University, Ankara, Turkey}
\author{A. Ozpineci}
\affiliation{Department of Physics, Middle East Technical University, Ankara, Turkey}
\author{D. Yildirim Yilmaz}
\affiliation{Department of Physics, Faculty of Sciences, Ankara University, Ankara, Turkey }
\affiliation{Department of Physics, Faculty of Sciences and Arts, Amasya University, Amasya, Turkey}
\email{yildirimyilmaz@amasya.edu.tr}

\date{\today}

\begin{abstract}
In this study, the properties of the $J^{PC}=1^{++}$ hidden charm poles are analyzed under the variation of the bare 2P charmonium mass within the effective field theory proposed in Ref.~\cite{Cincioglu:2016fkm}. The main focus of the current work is on the pole trajectory of the $\chi _{c1}(2P)$ charmonium dressed by the $D \bar{D}^*$ meson loops. It is shown that the trajectories of the pole change radically for values of the bare charmonium mass above a certain value and also depending on how close the pole is to the threshold.
\end{abstract}

\maketitle
\section{Introduction}

In the previous decades, many states were found experimentally in the mass range of heavy hadrons that can not be easily explained within a quark model and a standard quarkonium picture \cite{Olsen:2014qna,Tanabashi:2018oca}. To understand the structure of these states is one of the important aims of hadron physics. For most of these states, only their masses are experimentally known, and information about their quantum numbers and decays modes are very scarce \cite{Olsen:2017bmm, Lebed:2016hpi}. Especially the content of $X(3872)$ among all the exotics has been the subject of various studies since its discovery. The mass of $X(3872)$, which is very close to the $D^0D^{*0}$ threshold \cite{Choi:2003ue, Acosta:2003zx, Abazov:2004kp, Aubert:2004ns, Chatrchyan:2013cld,  Aaij:2013zoa, Aaij:2020qga, Aaij:2020xjx}, and the isospin-breaking decay $X(3872) \rightarrow J/\psi \rho$, make it an ideal candidate for a $D\bar D^*$ hadronic molecule \cite{Chen:2016qju,  Guo:2017jvc}. Another important measurement that might give insight into the structure of X(3872) is its electromagnetic decay into $\psi(2S)$ and $J/\psi$. The ratio of the branching ratios of these decays is measured as \cite{Aubert_2009, Aaij_2014}:
\begin{equation}
\label{eq:d1}
R_{\psi\gamma}=\frac{B_{r}(X\to\psi(2S)\gamma)}{B_{r}(X\to J/ \psi \gamma)}=2.46\pm 0.64\pm 0.29 \, .
\end{equation}

This implies that, since the phase space for the decay $X \rightarrow \psi(2S)\gamma$ is much smaller than the phase space for the $X \rightarrow J/\psi \gamma$ decay, the amplitude for $X \rightarrow \psi(2S)\gamma$ is much larger. This is naturally expected in the quark model since in the quark model $X \rightarrow \psi(2S)\gamma$ is a $\Delta L=1$ transition. But, in the quark model, the predicted mass of the  $1^{++} (J^{PC})$ charm- anticharm state, called $\chi_{c1}(2P)$, is around $ 3.95$ GeV \cite{Godfrey:1985xj, Ebert:2011jc}, about 70 MeV higher  than the observed state. On the other hand, in Ref.~\cite{Guo:2014taa}, triangular $D D^{*} \bar{D}^{*}$ and simple $D\bar{D}^{*}$ loop contributions to the radiative amplitude were computed. It was concluded that the observed ratio allows the $X(3872)$ to be a hadronic molecule with the dominant component $D\bar{D}^{*}$. In Ref.~\cite{Cincioglu:2019gzd}, the effects of short-range contributions to the radiative decays of the $X(3872)$ were analyzed. It was demonstrated that the possible constructive or destructive interferences between the meson-loop and the short-distance contact term are important to determine whether the charmonium content of the $X(3872)$ is nontrivial.

In \cite{Cincioglu:2016fkm}, an effective theory using heavy quark spin symmetry (HQSS) is presented in which $X(3872)$ is described as a superposition of molecular components and a compact core, taken to be 2P charmonium state throughout this study\footnote{ There are plenty of studies that considered X(3872) as a mixture in the literature \cite{ Dong:2009uf, Takizawa:2012hy, Chen:2013upa}. }. An important source of uncertainty in the model is bare masses of the compact components. In \cite{Cincioglu:2016fkm}, a bare mass was taken to be the charmonium mass predicted by potential quark models. This work aims to draw attention to the effect of the bare charmonium mass, which affects the trajectories on the predictions of the model, and, as shown, how it alters the molecular weight of the observation. It is noted that the bare mass is UV regulator dependent and it is a free parameter in the presented scheme. Although the bare mass is not physically observable, theoretically it is still relevant, because its value can be obtained from schemes that ignore the coupling of charmonium states to the mesons ($d=0$).  A major problem is to set the UV regulator to match the quark model and the EFT approaches. Thus, all calculations have been performed with two different UV cutoffs, spanning a physically motivated range of values. The expectation is that the cutoff dependence will be absorbed into the low energy constants (LECs) and thus predictions for observables could become at most mildly regulator dependent.

This paper is organized as follows: In section II, the main points of the model proposed in \cite{Cincioglu:2016fkm} is presented, followed by the our results and discussion.

\section{Formalism}

Due to the presence of heavy quarks in $X(3872)$, it can be described within a HQSS framework. To describe the molecular component of $X(3872)$, interactions of $D$ and $D^*$ are needed.  In the HQSS, these mesons group in a HQSS doublet, which can be written as:
\begin{equation}
H^{(Q)}_a = \frac{1+\slashed{v}}{2} (P^{*(Q)} _{a \mu} \gamma ^{\mu} - P^{(Q)}_a \gamma _5) \, .
 \label{d2}
\end{equation}
Due to the very low momentum exchange between the mesons in the molecule, contact interactions are sufficient to describe the D mesons' interactions in X(3872) \cite{Nieves:2012tt}.  Except for contact interaction, other interactions like one pion exchange and coupled channel are sub-leading order \cite{Nieves:2012tt}. Therefore their contribution can be ignored safely.  Since the interaction among the heavy hadrons forming a molecule is nonperturbative,  the potential should be iterated by solving Lippmann-Schwinger equation.  Lippmann-Schwinger equation shows ill-defined ultraviolet behavior resulting from a contact interaction. In consequence, it requires regularization.  As a regulator function, a Gaussian function $f_{\Lambda}(\vec{p})$ is employed
\begin{equation}
\langle \vec{p}^{\prime} ; D^{(*)}\bar{D}^{(*)} \vert V_{\Lambda} \vert \vec{p};D^{(*)} \bar{D}^{(*)} \rangle=C_{0X} f_{\Lambda}(\vec{p}^{\prime}) f_{\Lambda}(\vec{p}) 
\end{equation}
\begin{equation}
\langle \vec{p}; D^{(*)}\bar{D}^{(*)} \vert V_{c\bar{c}; \Lambda} \vert \Psi_{c\bar{c}}(2P) \rangle = d  f_{\Lambda}(\vec{p}) 
\end{equation}
where $d$ and $C_{0X}$ are low energy constants of related interactions,  and in this paper we take  $\Lambda$  cutoff values a $0.5-1.0$ GeV.

At leading order the interaction of four heavy mesons with contact interaction potentials can be described as below \cite{Nieves:2012tt}
\begin{equation}
 \begin{array}{l l l l l}
\mathcal{L}_{4H}&=&D_{0a} Tr\left[ \bar{H}^{(Q)a} H^{(Q)}_a \gamma_\mu \right] Tr\left[  H^{(\bar{Q})b} \bar{H}^{(\bar{Q})}_b \gamma^\mu \right]  \\
& + & D_{0b} Tr\left[ \bar{H}^{(Q)a} H^{(Q)}_a \gamma_\mu \gamma_5\right] Tr\left[  H^{(\bar{Q})b} \bar{H}^{(\bar{Q})}_b \gamma^\mu \gamma_5\right]    \\
&+&  E_{0a} Tr\left[  \bar{H}^{(Q)a}\vec{\tau}^b_a  H^{(Q)}_b \gamma_\mu \right] Tr\left[  H^{(\bar{Q})r} \vec{\tau}^s_r \bar{H}^{(\bar{Q})}_s \gamma^\mu \right]   \\
&+&E_{0b} Tr\left[ \bar{H}^{(Q)a}\vec{\tau}^b_a  H^{(Q)}_b \gamma_\mu \gamma_5\right] Tr\left[  H^{(\bar{Q})r} \vec{\tau}^s_r \bar{H}^{(\bar{Q})}_s \gamma^\mu \gamma_5\right] \, ,
 \label{d3}
\end{array}
\end{equation}
where $D_{0i} $ and $E_{0i} $ are LECs. To include the compact core component, it is necessary to identify the HQSS multiplet that can be used to describe it. For this purpose, the $P$-wave quarkonium multiplet, which can be written as \cite{Casalbuoni:1992yd}:
\begin{equation}
J^{\mu} = \frac{ 1+\slashed{v} }{2} \left( \chi_2 ^{\mu \alpha} \gamma _{\alpha} +\frac{ i }{\sqrt{2}} \epsilon ^{ \mu \alpha \beta \gamma} \chi _{1 \gamma} v_{\alpha} \gamma _{\beta} + \frac{1}{\sqrt{3}} \chi _0 (\gamma ^{\mu}-v ^{\mu}) + h ^{\mu} \gamma _5\right) \frac{1-\slashed{v}}{2} ~.
\label{d4}
\end{equation}
HQSS restricts the possible contact interactions between the $J^\mu$ multiplet, and the $D$ meson multiplets. The only interaction term between two $D$ mesons and $J^\mu$ that does not contain any derivative interactions at the leading order, and the only consistent Lagrangian with HQSS is:
\begin{equation}
\mathcal{L}_{HHQ\bar{Q}}=\frac{d}{2}Tr[H^{a(\bar{Q})}\bar{J}_{\mu} H_a^{(Q)}\gamma^{\mu}]+\frac{d}{2}Tr[\bar{H}^{a(Q)}J_{\mu} \bar{H}_a^{(\bar{Q})}\gamma^{\mu}] \, ,
\label{d5} 
\end{equation}
where the parameter $d$ is an unknown LEC that causes molecular and compact components to mix (For more details, see e.g. \cite{Cincioglu:2016fkm, Hanhart:2014ssa, Colangelo:2003sa}).

For bound states, to study the weights of the molecular and compact components in $X(3872)$, a method put forward by Weinberg \cite{Weinberg:1962hj, Weinberg:1965zz}  can be used. The method is crucial to examine interaction couplings and the probabilistic interpretations of the components. For small binding energies (s-wave), the approach is model-independent. With the help of the sum rule \cite{Garcia-Recio:2015jsa, Weinberg:1965zz}
\begin{equation}
-1=\sum_{ij}g_i g_j \left(\delta_{ij}\left[\frac{\partial G_i ^{II}(E)}{\partial E}\right]_{E=E_R} + \left[G_i ^{II}(E)\frac{\partial V_{ij}(E)}{\partial E} G_j ^{II}(E)\right]_{E=E_R} \right)   \, ,
 \label{d6}
\end{equation} 
the probabilistic interpretation of the compositeness condition can be made; moreover, Eq.~(\ref{d6}) is valid for both bound states and resonance states\cite{Garcia-Recio:2015jsa}. For resonance (bound) states, $G$ should be taken as $G^{II}$ ($G^I$). Each term in Eq.(\ref{d6}) can be identified differently\footnote{ The imaginary parts of $\tilde{X}_i$ and $\tilde{Z}$ must cancel each other.}.
\begin{equation}
X_i =Re\tilde{X}_i =Re \left(- g_i^2 \left[\frac{\partial G_i^{II}(E)}{\partial E}\right]_{E=E_R} \right) 
 \label{d7}
\end{equation}
\begin{equation}
Z =Re\tilde{Z}_i =Re \left(-\sum_{ij} \left[ g_i   G_i^{II}(E) \frac{\partial V_{ij}(E)}{\partial E} G_j^{II}(E)g_j\right]_{E=E_R}   \right)
\label{d8}
\end{equation}

With the definitions in Eqs.(\ref{d7}) and (\ref{d8}), we obtain the compositeness and elementariness, respectively. $\tilde{X}_i$  quantifies the probability of finding a two-body component in the wave function of a hadron, and $\tilde{Z}_i$ is related to other components and thus is understood as the elementariness. Hence Z close to 1 signifies that its compact component dominates the bound state.

On the other hand, in the case of a resonance, probabilistic interpretations are not entirely accurate because of $\tilde{X}_i$'s negative imaginary values. But in Ref.~\cite{Guo:2015daa, Aceti:2014ala}, it was claimed that the absolute value of $\tilde{X}_i$ can be used as a measure of the weight of the i-th channel\footnote{In Ref.~\cite{Guo:2015daa} a probabilistic interpretation of the compositeness relation at the pole of a resonance with only positive coefficients thanks to a suitable transformation of the S matrix has been derived. Absolute value of $\tilde Z_{i}$ gives the weight of finding a specific component in the wave function of a hadron, but it is only valid when $Re(E_R)>M_{i,th}$, with $E_R$ resonance energy and $M_{i,th}$ the corresponding threshold of the channel $i$.}.

$T$-matrix, which can be obtained as a solution of an LSE equation, develops poles on the complex energy plane. If $T$-matrix is close to the pole, its elements are approximately
\begin{equation}
T_{ij} \approx \frac{g_i g_j}{E-E_R}  \, ,
 \label{d9}  
\end{equation}
where $g_i$ is the coupling of the state to the i-th channel. 

When considering a specific $(1^{++})$ state, $T$-matrix, which gives dynamics of the system, can be formed as \cite{Cincioglu:2016fkm}:
\begin{equation}
T(E) = \frac{\Sigma_{c\bar{c}} }{1 - G^0_{c\bar{c}} \Sigma_{c\bar{c}}}    \left( \begin{array}{cc}
f^2_{\Lambda}(E)[\frac{1}{d^2 G_{QM}^2} - \frac{1-G^0 _{c\bar{c}} \Sigma_{c\bar{c}}}{G_{QM} \Sigma_{c\bar{c}}}] & f_{\Lambda}(E) \frac{1}{d G_{QM}}  \\
 f_{\Lambda}(E) \frac{1}{d G_{QM}} & 1  \end{array} \right) \, ,
 \label{d10}  
\end{equation}
where  $\Sigma_{c\bar{c}}$, $ G^0_{c\bar{c}}$, $ f_{\Lambda}$\footnote{ All calculations have been performed with an UV cutoff $\Lambda=0.5-1$ GeV. } , and $ G_{QM}$ \footnote{ QM stands for non-relativistic quantum mechanics.} are the charmonium self-energy induced by the meson loops, the non-relativistic bare charmonium propagator, the Gaussian regulator, and the diagonal meson loop function, respectively. In Eq.~(\ref{d10}), while the first channel is molecular type, the second is charmonium \cite{Baru:2010ww}. Besides, poles of the transition matrix are given by zeros of the inverse of the dressed propagator \cite{Cincioglu:2016fkm}
\begin{equation}
 1 - G^0 _{c\bar{c}}(E_R) \Sigma _{c\bar{c}}(E_R) = 0  \, ,
 \label{d11}  
\end{equation}
where $\Sigma_{c\bar c}$ is the quarkonium self-energy 
\begin{equation}
\Sigma_{c\bar c}(E) = \left[V_{c\bar c}^{\rm QM}\right]^t G_{\rm QM}(E)\Gamma_{c\bar c}(E) \, ,
 \label{d12}  
\end{equation}
with the dressed vertex function, $\Gamma_{c\bar c}$, reads
\begin{equation}
\Gamma_{c\bar c}(E) = \left( 1-V^{\rm QM} G_{\rm QM}(E)\right)^{-1}V_{c\bar c}^{\rm QM} \, .
 \label{d13}  
\end{equation}
where $V^{QM}$ and $V_{c\bar c}^{QM}$ are the molecular contact potential and the $\chi_{c_1}(2P)-D\bar{D}^{(*)}$ transition amplitudes, respectively (see Eq.~(30) and Eq.~(34) of Ref.~\cite{Cincioglu:2016fkm}). Finally,  for an arbitrary $E$, the mesonic loop function is given by~\cite{Albaladejo:2013aka}
\begin{align}
G_{\rm QM}(E) & = 
\int \frac{\text{d}^3 \vec{q}}{(2\pi)^3} \frac{e^{-2\vec{q}^{\,2}/\Lambda^2}}{E-M_1-M_2 - \vec{q}^{\,\,2}/2\mu + i0^+} \nonumber\\
& = -\frac{\mu\Lambda}{(2\pi)^{3/2}} + \frac{\mu k}{\pi^{3/2}}\phi\left(\sqrt{2}k/\Lambda\right)-i \frac{\mu k}{2\pi}e^{-2k^2/\Lambda^2}~,\label{eq:gmat_gr}
\end{align}
with $\mu^{-1}=M_1^{-1}+M_2^{-1}$, $k^2= 2\mu (E-M_1-M_2)$ and $\phi(x)$ the Dawson integral given by:
\begin{equation}
\phi(x) =  e^{-x^2}\int_{0}^{x} e^{y^2} \text{d}y~.
\end{equation}
Poles on the complex E-energy plane represent the observable states. The mass and width of a pole can be obtained from the pole position on the complex energy plane. On the complex energy E-plane, poles can be located on the different Riemann sheets. Indeed, $G_{\rm QM}(E)$ has two Riemann sheets. In the first Riemann sheet (FRS), $0\leqslant{\rm Arg}(E-M_1-M_2)< 2\pi$, there is a discontinuity $G_{\rm QM}^I(E+ i\epsilon)-G_{\rm QM}^I(E-i \epsilon) = 2i\,{\rm Im}G_{\rm QM}^I(E+i\epsilon)$ for $E> (M_1+M_2)$. For those poles located on the FRS, on the real axis, and below the threshold are named bound states. In the second Riemann sheet (SRS), $2\pi\leqslant {\rm Arg}(E-M_1-M_2)< 4\pi$, one can find $G_{\rm QM}^{II}(E- i\epsilon) = G_{\rm QM}^I(E+i\epsilon)$, for real energies and above threshold. Poles located below the real axis, and above the threshold on the SRS are called resonances\footnote{ The more detailed information about poles can be found in Refs.~\cite{Guo:2017jvc,  Hanhart:2014ssa}. There is no restriction for the location of the poles on the second Riemann (unphysical) sheets. Hermitian analyticity requires that if there is a pole at a complex value of s (resonance), there must be another pole at its complex conjugate value,  $s^*$ (anti-resonance). In this study,  the properties of the conjugate pole are not given since this pole corresponds to the same resonance.}. 

When it is a narrow resonance in SRS, the pole with a negative imaginary part (the pole located in the lower half-plane) is closer to the physical Riemann sheet than the pole with a positive imaginary part; thus, it influences the observables more strongly in the vicinity of the resonance region \cite{Baru:2010ww}. Moreover, when the poles' real part reaches the threshold with increasing d, both resonance and anti-resonance poles are equally essential. Those nearby poles only significantly influence resonance behavior in the experiment region that could be extracted from the experiment data in a phenomenological study. However, when they have large imaginary parts, they lose their width interpretation. 

The position of the pole might give further insight into the structure of the state. It appears that if the bound state is mostly compact, there are two near-threshold poles, one is on the first Riemann sheet, and the other is on the second Riemann sheet. Furthermore, if it is a predominantly molecular state, there is a single near-threshold pole on the first Riemann sheet \cite{Baru:2010ww, Morgan:1992ge}.

We look carefully at the trajectories of 2P charmonium poles located in nearby threshold zone in scattering amplitudes. There are qualitative differences between the pole trajectories of resonances that couple to the related continuum channel with changing bare 2P charmonium masses.
%

\subsection{General remarks}

In the model of Ref.~\cite{Cincioglu:2016fkm}, two poles in the $1^{++}$ sector are expected. One is located in the FRS as a bound state. This state is identified as $X(3872)$ bound state in the FRS, and its mass is fixed at $3871.69$ MeV to determine the LEC $C_{0X}$ defined as
\begin{equation}
C_{0X}= C_{0A}+C_{0B}, \qquad C_{0\phi}= D_{0\phi}+3 E_{0\phi}, \qquad  \text{for} \qquad \phi=a,\, b.
\end{equation}
The other one is identified as a dressed $\chi _{c1}(2P)$. The dependence of the second pole position on the bare 2P charmonium mass results from the non-relativistic bare propagator:
\begin{equation}
G^0_{c\bar{c}}(E)= \frac{1}{E-\mbare} \, ,
 \label{d14}
\end{equation} 
where $\mbare$ is the mass of the  2P bare charmonium state. As mentioned above, it is dressed by the $D\bar{D}^{*}$ meson loops and gives rise to the physical mass of the charmonium states, as $d$ increases. 

On the other hand, there exist some uncertainties in the presented model of Ref.~\cite{Cincioglu:2016fkm}. One of the most considerable uncertainties is the mass of the bare charmonium state. As can be seen in Table~\ref{tab:1}, most recent constituent quark models give the mass of the $\chi_{c1}(2P)$ in quite a broad range\footnote{ Furthermore, if the compact component is identified as a tetraquark, its mass is completely unknown.}. In  Ref.~\cite{Cincioglu:2016fkm}, the bare $\chi_{c1}(2P)$ mass is taken 3906 MeV from Ref.~\cite{Ebert:2011jc} due to the closest prediction to the experimental mass of the $\chi_{c2}(2P)$ state. Another uncertainty is a sizable error in the predicted masses that can reach up to $10\%$. Besides, different models in Table~\ref{tab:1} predict masses for the two-meson threshold and the bare  $\chi_{c1}(2P)$ mass differences ranging from around 35 MeV to 82 MeV \footnote{To see effects of the a lower mass of the threshold(for $3865$ MeV see Tables \ref{tab:01} and \ref{tab:05}), we searched the hypothetical values below the threshold. Contrary to the values higher than the threshold, as $C_{0X}$ decreases, $d$ increases, the pole moves towards the threshold, gaining small width.}. Also, it might be interesting to see the impact of the bare charmonium mass, which is $10$ MeV below the $3906$ MeV mass value, on the properties of the $1^{++}$ hidden charm poles (charmonium content in $X(3872)$, dressed charmonium  $\chi_{c1}(2P)$, $DD^*- \chi_{c1}(2P)$ coupling, etc.). For $m_{c\bar{c}}^0=3906$ MeV, the state is just $\sim35$ MeV above the $DD^*$ threshold, and dressed  $\chi_{c1}(2P)$ pole becomes below threshold in the SRS with relatively small $X(3872)$ charmonium content. However, as the larger bare $\chi_{c1}(2P)$ mass is taken, it is seen that on the SRS, a larger charmonium content is needed to move the $\chi_{c1}(2P)$ state below the  $DD^*$ threshold. 
\begin{table}
\centering
\begin{tabular}{c|c|c}
 Ref. & $m_{\mathcal{X}_{c1}(2P)}$ [MeV] & $\mbare-m_{DD^*}$ [MeV]  \\[2pt]
\hline 
\hline  
\cite{Ebert:2011jc} &3906  & 35 \\[2pt] \hline
\cite{Sungu:2018eej}& 3924 & 53 \\ [2pt] \hline
\cite{Barnes:2005pb}& 3925 & 54 \\ [2pt] \hline
\cite{Ebert:2002pp}& 3929 & 58  \\ [2pt] \hline
\cite{Gui:2018rvv, Deng:2016stx}&3937& 66 \\ [2pt] \hline
\cite{Segovia:2013wma, Ortega:2012rs}& 3947 & 76  \\[2pt] \hline
\cite{Godfrey:1985xj}& 3953 & 82  \\ [2pt]
\hline \hline
\end{tabular} 
\caption{The $\chi_{c1}(2P)$ masses in the literature. Constituent quark model~\cite{Segovia:2013wma, Ortega:2012rs}, Regge trajectory~\cite{Ebert:2011jc}, relativistic quark model~\cite{Ebert:2002pp}, non-relativistic potential model~\cite{Barnes:2005pb, Godfrey:1985xj}, non-relativistic quark model - $^3P_0$ model~\cite{Gui:2018rvv, Deng:2016stx}, and QCD sum rule~\cite{Sungu:2018eej} were used to obtain these mass values.}
\label{tab:1}
\end{table}
%


\section{Results and Discussion}

In Tables~\ref{tab:2}--\ref{tab:8} and Table I of Ref.~\cite{Cincioglu:2016fkm}\footnote{ Table I includes properties of $\chi_{c1}(2P)$ states, taking the bare charmonium mass as 3906 MeV. }, the properties of the poles found in the $1^{++}$ hidden charm sector are studied as a function of the LEC $d$, which controls the admixtures of charmonium and $DD^*$ molecule. The location of the dressed  $\chi_{c1}(2P)$ pole depends on the mixing parameter $d$ and the bare charmonium mass. Within the scheme presented here and in Ref.~\cite{Cincioglu:2016fkm} of bare charmonium mass is a free parameter and it is not an observable, as mentioned above, which gets dressed by the $D^{(*)}D^{(*)}$ meson loops and gives rise to the physical mass of the charmonium states. Couplings of the $\chi_{c1}(2P)$ state to $D^{(*)}$ mesons causes the bare mass to be renormalised. Since, in the effective theory, the difference between the bare and the physical charmonium masses is a finite renormalization. This shift depends on the UV regulator since the bare mass itself depends on the renormalization scheme. To show the cutoff dependency of the properties of the poles found in the $1^{++}$ hidden charm sector, all calculation have been carried out with UV cutoff values $\Lambda=0.5\--1$ GeV. The results obtained for both cutoffs with values of d are qualitatively similar, though some quantitative differences appear, as can be seen in Tables~\ref{tab:2}--\ref{tab:8} and Tables~\ref{tab:9}--\ref{tab:14} of the appendix. Moreover, it is observed that the behavior of the physical $\chi_{c1}(2P)$ state alters around $\mbare \approx 3908.5$ MeV.  

\begin{center}
\begin{figure}[h]
\includegraphics[width=1.1\textwidth]{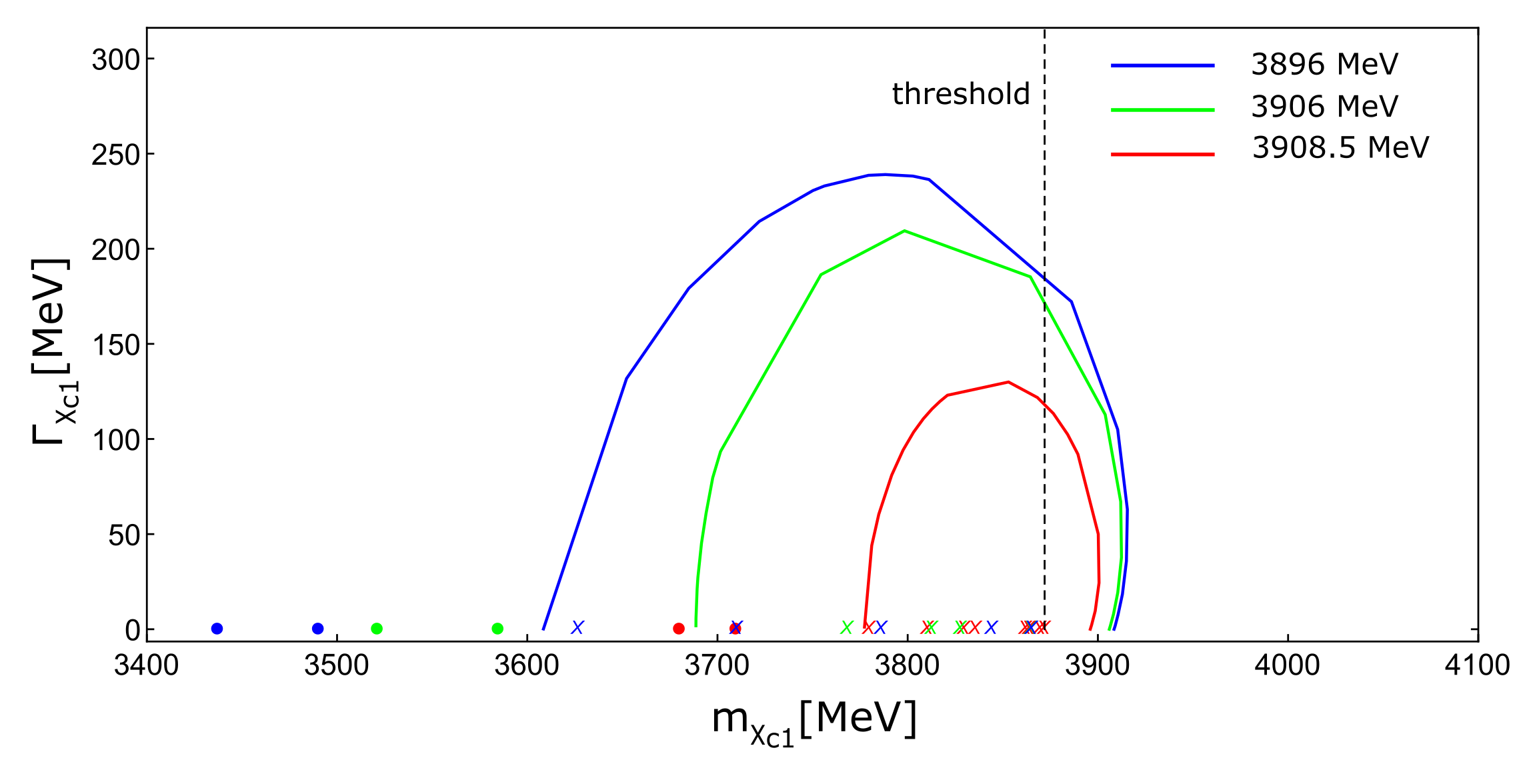}
\caption{\small{ Pole trajectories of the $\chi_{c1}(2P)$, located in the SRS, for the different bare charmonium masses for $\mbare=3896, \, 3906$ and 3908.5 MeV. The $D\bar{D}^{*}$ threshold is shown as a vertical black dashed line. The lines and the crosses are obtained with the help of Tables~\ref{tab:2}, \ref{tab:3} and Table I of Ref. ~\cite{Cincioglu:2016fkm}.  While the crosses show the trajectories which come close to the threshold after colliding the real axis, and the solid circles show the trajectories which move away from the threshold. Note that the properties of the poles that depart from the threshold are not given in Tables~\ref{tab:2}, \ref{tab:3}.} 
\label{fig:1}}
\end{figure} 
\end{center}

For $\mbare = 3896, \, 3906$ and $3908.5$ MeV values, the trajectories of the corresponding dressed $\chi_{c1}(2P)$ pole are depicted in Fig.~\ref{fig:1}. When $d$ is $0$, the pole in the SRS is on the real axis. With the increasing value of $d$, the pole gradually gets away from the real axis by gaining width. As d continues to increase, the pole moves below the threshold. At some point, it reaches the real axis again. When the pole reaches the real axis with $\tilde X_{\X} \sim 0.39$ for $\mbare =3896$ MeV,  $\tilde X_{\X} \sim 0.43$ for $\mbare = 3906$ MeV and $\tilde X_{\X} \sim 0.47$ for $\mbare = 3908.5$ MeV.  As can be seen at Tables~\ref{tab:2}--\ref{tab:3}, for  $m_{c\bar{c}}^0=3896$ MeV charmonium content $\vert \tilde{X}_{\chi_{c1}} \vert $ is about 35$\%$, and for  $m_{c\bar{c}}^0=3908.5$ MeV charmonium content $\vert \tilde{X}_{\chi_{c1}} \vert $ is about 83$\%$. With the conjugate pairs coinciding in the real axis, two poles appear in the SRS below threshold located at $m_R-0i$. As one pole moves along the real axis toward the threshold, a second pole departs from the threshold and leaves the real axis forming another conjugate pair, with $d$ increasing. These new-formed poles are either far below the threshold or above the threshold but deep in the complex plane. It means that, in that region, the width interpretation is lost. Since it will not produce any observable effects, its behavior is not illustrated in Fig.~\ref{fig:1}, and also its properties are not given in Tables~\ref{tab:2} and \ref{tab:3}. Additionally, values smaller than $3908.5$ MeV bare charmonium mass have similar trajectories.

\begin{center}
\begin{figure}[h]
\includegraphics[width=1.1\linewidth]{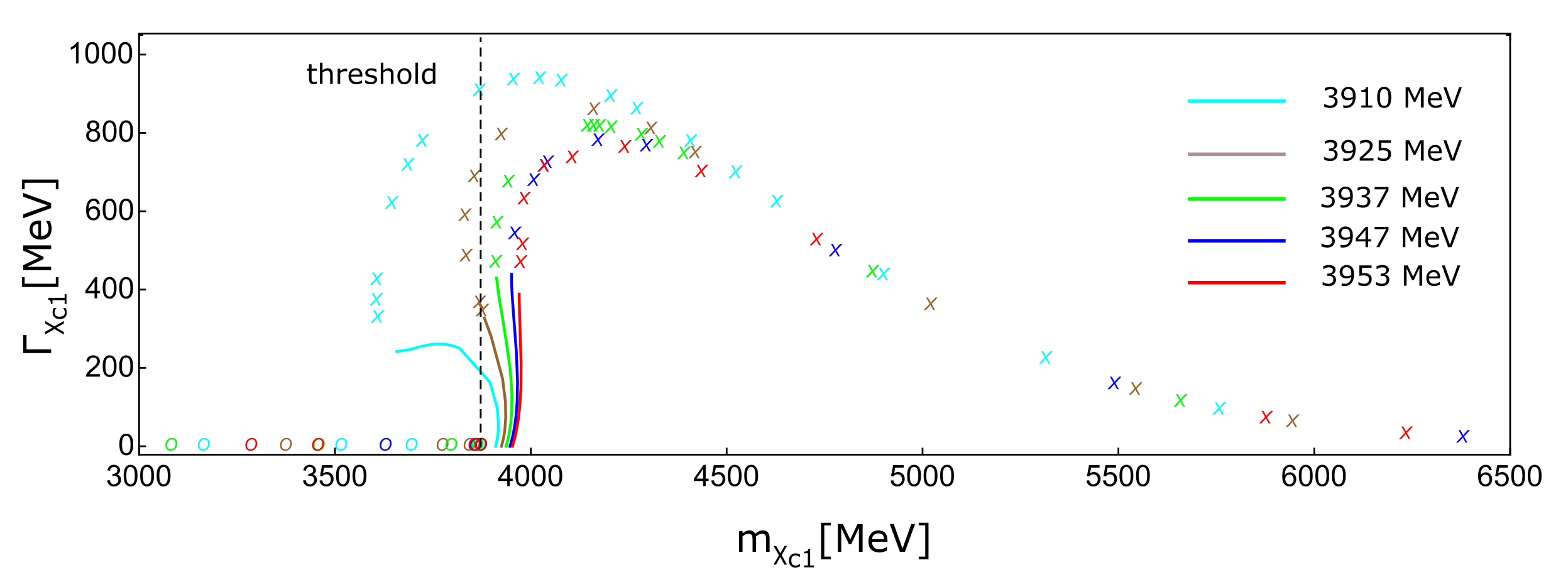}
\caption{\small{ Pole trajectories of the $\chi_{c1}(2P)$, located in the SRS,  for the different bare charmonium masses for $\mbare=3910, \, 3925, \, 3937, \,3947$ and 3953 MeV. The $D\bar{D}^{*}$ threshold is shown as a vertical black dashed line. To present a general picture, those located in the much above the real axis do not have any observable effects and are illustrated with crosses. The circles show other pole trajectories which come close to the threshold from deep. The lines and the circles are obtained with the help of the values in Tables~\ref{tab:4}--\ref{tab:8}. Note that the properties of the poles that depart from the threshold are not given in Tables~\ref{tab:4}--\ref{tab:8}.} 
 \label{fig:2}}
\end{figure}
\end{center}

In the case of $\mbare \gtrsim 3908.5$ MeV, the behaviors of the pole trajectories, depicted in Fig.~\ref{fig:2}, are different from Fig.~\ref{fig:1}. However, with increasing $d$ values, the poles stay in the SRS above or below the threshold. But they do not reach the real axis until d's high values. For larger bare mass values than $3908.5$ MeV, until around $3930$ MeV bare charmonium mass, the pole trajectories cross the threshold. For those who cross the threshold, while the bare charmonium mass is getting bigger, their molecular weight of the $X(3872)$ state is getting smaller. Besides this, for some specific values of $d$ where $ d> d^{\rm{crit}}$, there exist a double pole in the SRS.  One of the poles approaches the threshold, with $\Sigma_{c\bar c}^{\prime}$ decreasing and moves along the real axis in the SRS becomes quite close to the threshold, where SRS and FRS are connected,  it might have visible effects in scattering observables as the line shape will be determined by both the pole and this virtual state. The other goes away from the real axis with gaining width but eventually reaches the real axis far from the threshold. These trajectories of the poles (illustrated with crosses) as $d$ increases are either below the threshold or above the threshold but located in much above the real axis. As mentioned above, since these poles will not have any observable consequences, their details are not also included in Tables~\ref{tab:4}--\ref{tab:8} in the appendix. In the $d\gg d^{\rm{crit}}$ limit, $X(3872)$ appears to be a 2P charmonium state where the molecular weight in $X(3872)$, $X_{\X} \sim 0$, mirror in the FRS of the pole found in the SRS. Also, the proximity of the bare mass to the threshold is essential within the model.

As mentioned in the introduction, the radiative decays of $X(3872)$ were analyzed in Ref.~\cite{Cincioglu:2019gzd} within the effective theory of Ref.~\cite{Cincioglu:2016fkm} to constrain the charmonium content in $X(3872)$.  In the case of the destructive interferences between the meson loops and the counter-term modeled by a charm quark loop, a strong restriction on the charmonium admixture was found. In this work, the contribution from short range interaction to the ratio $R{\psi\gamma}$ depends on $\tilde{Z}_{X(3872)}$\footnote{In the presented study, it is defined as $\tilde{Z}_{X(3872)}=1-\tilde{X}_{X(3872)}$}, which is the weight of finding the charmonium component $\chi_{c1}(2P)$ in the physical wave function of $X(3872)$, and position of the $\chi_{c1}(2P)$ pole, as can be seen from Eq.~(4) of that reference. It was claimed that the behavior of the predictions of the ratio of radiative decays is different when $\tilde{Z}_{X(3872)}\lesssim 0.55$ and when $\tilde{Z}_{X(3872)}\gtrsim 0.55$\footnote{As can be seen from Fig.~2 of that reference there is a bump around $\tilde{Z}_{X(3872)}\sim$ 0.55.}. Indeed, in the vicinity of $C_{0X}$=0 ($\tilde{Z}_{X(3872)}\sim 0.55$), which controls the four-meson contact interaction, the dressed $\chi_{c1}(2P)$ pole becomes below threshold with its mass decreasing rapidly and quite wide. As $\tilde{Z}_{X(3872)}$ increases, the mass and width of the charmonium state decrease, up to $\tilde{Z}_{X(3872)} \sim 0.57$ ($\chi_{c1}(2P)$ pole reaches the real axis with that weight), while $C_{0X}$ increases and takes large positive values which creates a strong repulsive force between the $D$ and $D^*$ mesons. Thus, the contribution of the molecular component in the $X(3872)$ is suppressed. However, this mentioned behavior of radiative branching ratio is valid only for  $\mbare=3906$ MeV. For larger (smaller) values of bare charmonium, smaller (larger) $\chi_{c1}(2P)$ contents are required to reach the real axis of the pole. For instance, the $\chi_{c1}(2P)$ pole appears on the real axis below threshold when $\tilde{Z}_{X(3872)}\sim$ 0.61 for $\mbare=3896$ MeV and when $\tilde{Z}_{X(3872)}\sim$ 0.39 for $\mbare=3937$ MeV, as can be seen in Tables~\ref{tab:2}--\ref{tab:8} of the appendix. Due to these non-trivial effects, it is difficult to put a firm restriction on the charmonium admixture in $X(3872)$.

\subsection*{Acknowledgement}
This research has been supported by TUBITAK (The Scientific and Technological Research Council of Turkey) under the grant no F117090.
\clearpage
\section{Appendix}
\label{appendix}
In this appendix, for the values whose pole trajectories are shown with the Figs.~\ref{fig:1} and \ref{fig:2}, their tables are shown below.  The calculations have been carried out with an UV cutoff $\Lambda =0.5-1$ GeV. In the tables, $d^{crit}$ indicates the point that $C_{0X}$ is zero. After $d^{crit}$ values, $C_{0X}$ becomes positive, which means that the interaction becomes repulsive.  While details of the Fig.~\ref{fig:1} values are given by Tables~\ref{tab:2} and \ref{tab:3},  values in the Fig.~\ref{fig:2}  are given by Tables~\ref{tab:4}--\ref{tab:8}. Moreover, the properties of the $1^{++}$ hidden charm poles are compiled as a function of the mixing parameter $d$, when an UV cutoff is used to regularized the molecular interactions (see Tables~\ref{tab:9}--\ref{tab:14}). Note that $X(3872)$ is assumed as a bound state in the FRS; therefore,  the pole position of $X(3872)$ is fixed at 3871.69 MeV in the FRS. Finally, the position of the $\chi_{1c}(2P)$ is located in the SRS. 

Numerical results for an UV cutoff $\Lambda=1$ GeV:

\begin{table}[h]
\centering{
\scalebox{0.8}{ 
\begin{tabular}{c  |c  c  c  | c  c  c c}
d [fm$^{1/2}$] & C$_{0X}$ [$fm^{2}$] & $g_{D\bar{D}^*}^{X(3872)}$[GeV$^{-1/2}$] & $\tilde{X}_{X(3872)}$ & ($m_{\chi_{c1}},\Gamma_{\chi_{c1}} $)[MeV] & $g_{D\bar{D}^*}^{\chi_{c1}}$[GeV$^{-1/2}$] &$\vert \tilde{X}_{\chi_{c1}} \vert $  & $\tilde{Z}_{\chi_{c1}}$  \\ [2pt] \hline \hline
0.01  & -0.792 & 0.891 &0.978 & (3865.0, 0.1) &  0.02-0.05i &0.002  &1.00  \\ [2pt] 
0.05 & -0.862& 0.723 & 0.645 & (3866.1,  2.0)   &   0.11-0.20i    &  0.055&  1.05-0.01i \\[2pt]
0.10  & -1.084 & 0.503  &0.312  & (3868.7, 3.17)    &  0.25-0.20i  & 0.122  & 1.11+0.06i  \\[2pt]
0.20 & -1.968& 0.288 & 0.102 & (3870.8, 1.38)   & 0.21-0.07i   & 0.071   & 1.03+0.06i   \\[2pt]
0.40 & -5.508 & 0.150  & 0.028 &(3871.46,  0.39) & 0.11-0.03i  &  0.022& 1.01+0.02i \\[2pt]
1.00 & -30.285 & 0.061  &  0.005 & (3871.65,  0.06)& 0.05-0.01i   &0.004   & 1.00  \\[2pt]
3.00 & -266.251& 0.020  & 0.000   & (3871.69, 0.007) & 0.015-0.003i   & 0.00  &   1.00 \\[2pt]
10.00 & -2950.37& 0.006  & 0.000   & (3871.69, 0.000) & 0.005-0.001i   & 0.00  &   1.00 \\[2pt]

\hline \hline
\end{tabular} } }
\caption{\small{For $m^0_{c\bar{c}} = $ 3865 MeV bare charmonium mass, dressed mass value of the $\chi_{1c}(2P)$ and its other properties as a function of $d$. }} 
\label{tab:01}
\end{table}


\begin{table}[h]
\centering{
\scalebox{0.8}{ 
\begin{tabular}{c  |c  c  c  | c  c  c c}
d [fm$^{1/2}$] & C$_{0X}$ [$fm^{2}$] & $g_{D\bar{D}^*}^{X(3872)}$[GeV$^{-1/2}$] & $\tilde{X}_{X(3872)}$ & ($m_{\chi_{c1}},\Gamma_{\chi_{c1}} $)[MeV] & $g_{D\bar{D}^*}^{\chi_{c1}}$[GeV$^{-1/2}$] &$\vert \tilde{X}_{\chi_{c1}} \vert $  & $\tilde{Z}_{\chi_{c1}}$  \\ [2pt] \hline \hline
0.00  & -0.789 & 1.00 &1.00 & (3896.0, 0.0) &   0.00 &0.00  &1.00  \\ [2pt] 
0.05 & -0.768 & 0.88 & 0.96 & (3896.7, 2.2)   &  0.01-0.19i   &  0.03 &  0.98+0.02i     \\[2pt]
0.10  & -0.707 & 0.83  &0.86  & (3898.6, 9.6)    &  0.00-0.36i   & 0.10  & 0.93+0.07i    \\[2pt]
0.20 & -0.464& 0.70  & 0.60 & (3900.2, 50.0)   & 0.16+0.63i    & 0.35   & 0.80+0.29i    \\[2pt]
0.30 & -0.058 & 0.57 & 0.40 & (3821.0, 123.1)   & 0.82+1.02i    &  $>$1  &  0.64+1.79i   \\[2pt]
0.305 & -0.034 &  0.56 & 0.39  & (3797.6,  94.2)   &  1.10+1.25i   &   $>$1  &  0.62+3.02i   \\[2pt]
0.307 & -0.024 &  0.56 & 0.39 & (3784.8, 60.5)   &  1.48+1.60i   &   $>$1  &   0.60+5.33i  \\[2pt]
0.3075& -0.021&  0.56 & 0.39 &(3781.1, 43.9) &  1.79+1.88i   &  $>$1   &  0.59+7.59i  \\[2pt]
0.3078& -0.019&  0.56 & 0.39 &(3778.8, 28.1) &  2.28+2.35i   &  $>$1   &  0.59+12.12i  \\[2pt]
0.30798& -0.019&  0.56 & 0.39 &(3777.4, 7.9) &  4.35+4.39i   &  $>$1   &  0.59+43.38i  \\[2pt]
0.309& -0.014&  0.56 & 0.39 &(3734.2, 0.0) &  0.00+2.25i   &  $>$1   &  -4.84  \\[2pt]
0.31& -0.008&  0.56 & 0.38 &(3810.7, 0.0) &  1.73  & $\vert \tilde{X}_{\chi_{c1}} \vert <1 $  & 4.44  \\[2pt]
$\emph{d}^{\emph{crit}}$ & 0.000 & 0.55 &0.38 & (3818.8, 0.0) &  1.45  & $\vert \tilde{X}_{\chi_{c1}} \vert <1 $& 3.45     \\[2pt]
0.35 & 0.206 & 0.52  & 0.33  &(3853.3, 0.0) &  0.66  & $\vert \tilde{X}_{\chi_{c1}} \vert <1 $& 1.57  \\[2pt]
0.40 & 0.510 & 0.47  & 0.27 &(3861.6, 0.0) & 0.47  &   $\vert \tilde{X}_{\chi_{c1}} \vert <1 $& 1.33 \\[2pt]
0.50 & 1.241  & 0.40 & 0.19 &(3866.8, 0.0) & 0.33  & $\vert \tilde{X}_{\chi_{c1}} \vert <1 $ & 1.18  \\[2pt]
1.00 & 7.328  & 0.21  &  0.06& (3870.8, 0.0)&0.14   &$\vert \tilde{X}_{\chi_{c1}} \vert <1 $ & 1.04   \\[2pt]
1.50 & 17.475  & 0.15  & 0.03 & (3871.3, 0.0)&  0.10  &  $\vert \tilde{X}_{\chi_{c1}} \vert <1$&  1.02   \\[2pt]
2.00 & 31.680  & 0.11  & 0.02   & (3871.5, 0.0) & 0.07   & $\vert \tilde{X}_{\chi_{c1}} \vert <1 $  &   1.01  \\[2pt]
3.00 & 72.265 &0.07&  0.01  &(3871.6, 0.0) & 0.05  &  0.00   &  1.00  \\[2pt]
\hline \hline
\end{tabular} } }
\caption{\small{For $m^0_{c\bar{c}} = $ 3896 MeV bare charmonium mass, dressed mass value of the $\chi_{1c}(2P)$ and its other properties as a function of $d$ ($ d^{crit}=0.311708 $ fm$^{1/2}$).  }} 
\label{tab:2}
\end{table}

\begin{table}
\centering{
\scalebox{0.8}{ 
\begin{tabular}{c  |c  c  c  | c  c  c c}
d [fm$^{1/2}$] & C$_{0X}$ [$fm^{2}$] & $g_{D\bar{D}^*}^{X(3872)}$[GeV$^{-1/2}$] & $\tilde{X}_{X(3872)}$ & ($m_{\chi_{c1}},\Gamma_{\chi_{c1}} $)[MeV] & $g_{D\bar{D}^*}^{\chi_{c1}}$[GeV$^{-1/2}$] &$\vert \tilde{X}_{\chi_{c1}} \vert $  & $\tilde{Z}_{\chi_{c1}}$  \\ [2pt] \hline \hline
0.00  & -0.789 & 1.00  &1.00 & (3908.0, 0.0) &   0.00  &  0.00 & 1.00  \\ [2pt] 
0.05 & -0.775 & 0.89 & 0.98 & (3909.6, 1.8)   &   0.01+0.15i  &  0.01   &0.99+0.01i    \\[2pt]
0.10  & -0.735 & 0.87  &  0.93 & (3910.6, 7.6)    & 0.03+0.30i  & 0.06  &  0.97+0.05i   \\[2pt]
0.20 & -0.574 & 0.79  & 0.77 & (3915.0, 35.8)   & 0.14+0.54i   & 0.21   & 0.89+0.18i    \\[2pt]
0.30 & -0.306&  0.70 & 0.60 & (3910.4, 104.9)   &  0.35+0.71i   &   0.49 & 0.79+0.44i    \\[2pt]
0.35 & -0.132 & 0.65  &0.53   &(3886.1, 172.1)    &  0.55+0.80i   &   0.83  & 0.73+0.78i    \\[2pt]
0.38 & -0.014 & 0.63  &0.49   &(3827.9, 230.5)    &  0.80+0.97   &   $>$1  & 0.60+1.55i    \\[2pt]
$\emph{d}^{\emph{crit}}$  & 0.000  & 0.63 & 0.48&(3811.2, 236.5) &0.87+1.04i  &$>$1    &  0.54+1.84i   \\[2pt]
0.39 & 0.027 & 0.62  & 0.47 &(3756.2, 233.0)  & 1.13+1.34i  & $>$1  & 0.16+3.31i    \\[2pt]
0.391 & 0.031 & 0.62  & 0.47 &(3739.8, 225.7)  & 1.24+1.47i  & $>$1  & -0.05+4.02i    \\[2pt]
0.393& 0.039& 0.62  & 0.47 & (3672.0, 162.6) & 1.98+2.58i  & $>$1 & -3.05+12.07i    \\[2pt]
0.3932& 0.040 & 0.62  & 0.47 & (3652.2, 131.8) & 2.49+3.39i  & $>$1 & -6.48+20.28i    \\[2pt]
0.39334& 0.041& 0.62  & 0.47 & (3616.6, 49.3) & 6.50+9.19i  & $>$1 & -54.59+148.92i    \\[2pt]
0.393345& 0.041& 0.62  & 0.47 & (3611.2, 27.1) & 10.84+14.09  & $>$1 & -104.57+383.37i    \\[2pt]
0.393346& 0.041& 0.62  & 0.47 & (3609.5, 16.3) & 15.72+18.77  & $>$1 & -136.33+742.74i     \\[2pt]
0.395& 0.048& 0.62  & 0.47 & (3741.6, 0.0) & 2.14  & $\vert \tilde{X}_{\chi_{c1}} \vert <1 $& 6.25  \\[2pt]
0.398& 0.061& 0.61  & 0.46 & (3774.3, 0.0) & 1.58  & $\vert \tilde{X}_{\chi_{c1}} \vert <1 $& 3.85  \\[2pt]
0.40 & 0.069& 0.61  &  0.46 &(3785.7, 0.0)  & 1.43    &  $\vert \tilde{X}_{\chi_{c1}} \vert <1 $  & 3.31    \\[2pt]
1.00 & 4.572  & 0.31  & 0.12 & (3869.4, 0.0) & 0.22  &  $\vert \tilde{X}_{\chi_{c1}} \vert <1 $ &  1.09  \\[2pt]
2.00 & 20.654  & 0.16  & 0.03 &(3871.2, 0.0) & 0.11 & $\vert \tilde{X}_{\chi_{c1}} \vert <1 $& 1.02     \\[2pt]
3.00 & 47.458  & 0.11  & 0.02 &(3871.5, 0.0) & 0.07 & $\vert \tilde{X}_{\chi_{c1}} \vert <1$ & 1.01     \\[2pt]\hline \hline

\end{tabular}  }}
\caption{ \small{ For $m^0_{c\bar{c}} =$  3908.5 MeV bare charmonium mass, dressed mass value of the $\chi_{1c}(2P)$ and its other properties as a function of $d$ ($ d^{crit}=0.383565$ fm$^{1/2}$).}} 
\label{tab:3}
\end{table}

\begin{table}
\centering{
\scalebox{0.9}{ 
\begin{tabular}{c  |c  c  c  | c  c  c c}
d [fm$^{1/2}$] & C$_{0X}$ [$fm^{2}$] & $g_{D\bar{D}^*}^{X(3872)}$[GeV$^{-1/2}$] & $\tilde{X}_{X(3872)}$ & ($m_{\chi_{c1}},\Gamma_{\chi_{c1}} $)[MeV] & $g_{D\bar{D}^*}^{\chi_{c1}}$[GeV$^{-1/2}$] &$\vert \tilde{X}_{\chi_{c1}} \vert $  & $\tilde{Z}_{\chi_{c1}}$  \\ [2pt] \hline \hline
0.00 & -0.789 & 1.00  & 1.00& (3910.0, 0.0) &   0.00 & 0.00  &1.00  \\ [2pt] 
0.05 & -0.776 & 0.89 &  0.98& (3910.6, 1.8)   & 0.01+0.15i    &0.01  &0.99+0.01i     \\[2pt]
0.10 & -0.737 & 0.87  & 0.94  & (3912.1, 7.5)    & 0.04+0.29i & 0.05    &  0.97+0.04i   \\[2pt]
0.20 & -0.583 & 0.80  &0.79  & (3916.6, 34.8)   &  0.14+0.53i &  0.20   & 0.89+0.17i    \\[2pt]
0.30 & -0.325&  0.71 & 0.62 & (3913.7, 100.7)   & 0.34+0.70i  & 0.46    &  0.80+0.42i   \\[2pt]
0.35 & -0.158 & 0.67  &0.55   &(3895.0, 164.3)    &  0.51+0.78i &0.74  & 0.75+0.70i    \\[2pt]
$\emph{d}^{\emph{crit}}$  & 0.000  & 0.63 &0.49 &(3818.6, 249.8)    &  0.85+1.01i & $>$1    & 0.54+1.72i    \\[2pt]
0.40 & 0.035& 0.63  & 0.48 &(3163.8, 0.0)  & 0.86 &  $\vert \tilde{X}_{\chi_{c1}} \vert <1 $  & 2.22  \\[2pt]
0.402 & 0.044& 0.62  & 0.48 &(3514.7, 0.0)& 3.30& $\vert \tilde{X}_{\chi_{c1}} \vert <1 $   & 15.71  \\[2pt]
0.40205 & 0.044& 0.62  & 0.48 &(3529.6, 0.0)&3.61& $\vert \tilde{X}_{\chi_{c1}} \vert <1 $   & 18.40 \\[2pt]
0.4021 & 0.044& 0.62  & 0.48 &(3545.8, 0.0)&3.97 & $\vert \tilde{X}_{\chi_{c1}} \vert <1 $  & 21.75 \\[2pt]
0.403 & 0.048& 0.62  & 0.48 &(3694.4, 0.0) & 2.82 &  $\vert \tilde{X}_{\chi_{c1}} \vert <1 $  & 0.52  \\[2pt]
0.403 & 0.050& 0.62  & 0.48 &(3714.3, 0.0) & 2.42 &  $\vert \tilde{X}_{\chi_{c1}} \vert <1 $  & 0.52  \\[2pt]
0.45 & 0.254 & 0.59  &0.42 & (3838.0, 0.0)& 0.78& $\vert \tilde{X}_{\chi_{c1}} \vert <1 $   &  1.73 \\[2pt]
0.50 & 0.499  & 0.55 & 0.37 &(3852.3, 0.0)& 0.59 & $\vert \tilde{X}_{\chi_{c1}} \vert <1 $  &  1.46\\[2pt]
1.00& 4.362  & 0.32 & 0.13 &(3869.2, 0.0)& 0.23 & $\vert \tilde{X}_{\chi_{c1}} \vert <1$  &  1.10 \\[2pt]
2.00 & 19.815  & 0.17 & 0.36 &(3871.1, 0.0)& 0.11 & $\vert \tilde{X}_{\chi_{c1}} \vert <1 $   &  1.03 \\[2pt]
3.00 & 45.569  & 0.11 & 0.02 &(3871.5, 0.0)& 0.07 & $\vert \tilde{X}_{\chi_{c1}} \vert <1 $   &  1.01 \\[2pt] \hline \hline

\end{tabular}  }}

\caption{ \small{For $m^0_{c\bar{c}} =$ 3910 MeV bare charmonium mass, dressed mass value of the $\chi_{1c}(2P)$ and its other properties as a function of $d$ ($ d^{crit}=0.391302$ fm$^{1/2}$).}}
 \label{tab:4}
\end{table}

\begin{table}
\centering{
\scalebox{0.8}{ 
\begin{tabular}{c  |c  c  c  | c  c  c c}
d [fm$^{1/2}$] & C$_{0X}$ [$fm^{2}$] & $g_{D\bar{D}^*}^{X(3872)}$[GeV$^{-1/2}$] & $\tilde{X}_{X(3872)}$ & ($m_{\chi_{c1}},\Gamma_{\chi_{c1}} $)[MeV] & $g_{D\bar{D}^*}^{\chi_{c1}}$[GeV$^{-1/2}$] &$\vert \tilde{X}_{\chi_{c1}} \vert $  & $\tilde{Z}_{\chi_{c1}}$  \\ [2pt] \hline \hline
0.00 & -0.789 & 1.00  & 1.00& (3925.0, 0.0) & 0.00 & 0.00     &  1.00   \\[2pt]
0.05 & -0.780 & 0.90 &  0.99& (3925.5, 1.5)  & 0.03+0.12i     & 0.01    &   1.00+00i  \\[2pt]
0.10 & -0.752 & 0.89  & 0.97  & (3926.8, 6.2) &  0.06+0.24i    &0.03    & 0.98+0.03i    \\[2pt]
0.20  & -0.641 & 0.84  & 0.88 & (3931.5, 27.5) &  0.15+0.45i    &0.13     &  0.94+0.11i   \\[2pt]
0.30 & -0.456&  0.79 & 0.76 & (3935.9, 73.3)  &  0.29+0.62i   &  0.29   &  0.87+0.26i   \\[2pt]
0.40 & -0.196& 0.72  & 0.64 &(3927.9, 172.3)   &  0.50+0.74i   & 0.59    & 0.79+0.56i    \\[2pt]
0.45 & -0.039 & 0.69  & 0.59& (3898.6, 280.6) & 0.69+0.84i    & $>$1    & 0.67+0.96i    \\[2pt]
$\emph{d}^{\emph{crit}}$  & 0.000& 0.68 &0.57 &(3882.1, 325.8)  &  0.76+0.90i   &  $>$1   & 0.58+1.18i    \\[2pt]
0.48 &  0.064 & 0.67 &0.56  &(3373.4, 0.0)& 0.99   & $\vert \tilde{X}_{\chi_{c1}} \vert <1 $ & 2.44  \\[2pt]
0.50 & 0.137  & 0.66 &0.54  &(3749.4, 0.0) & 1.20   & $\vert \tilde{X}_{\chi_{c1}} \vert <1 $  &  2.66  \\[2pt]
0.60 & 0.544  & 0.60 &  0.44& (3842.3, 0.0) & 0.65   &    $\vert \tilde{X}_{\chi_{c1}} \vert <1 $&1.53   \\[2pt]
1.00 & 2.913  & 0.43 &  0.22& (3866.2, 0.0) & 0.33   & $\vert \tilde{X}_{\chi_{c1}} \vert <1 $&1.18   \\[2pt]
2.00 & 14.017  & 0.23 &  0.07& (3870.6, 0.0) & 0.16   & $\vert \tilde{X}_{\chi_{c1}} \vert <1 $&1.05  \\[2pt]
4.00 &58.44  & 0.12 &  0.02& (3871.4, 0.0) & 0.08   & $\vert \tilde{X}_{\chi_{c1}} \vert <1$&1.01   \\[2pt] \hline \hline
\end{tabular}    }}
\caption{ \small{ For $m^0_{c\bar{c}} =$ 3925 MeV bare charmonium mass, dressed mass value of the $\chi_{1c}(2P)$ and its other properties as a function of $d$ ($ d^{crit}=0.461594$ fm$^{1/2}$).}}  \label{tab:5}
\end{table}

\begin{table}
\centering{
\scalebox{0.8}{ 
\begin{tabular}{c  |c  c  c  | c  c  c c}
d [fm$^{1/2}$] & C$_{0X}$ [$fm^{2}$] & $g_{D\bar{D}^*}^{X(3872)}$[GeV$^{-1/2}$] & $\tilde{X}_{X(3872)}$ & ($m_{\chi_{c1}},\Gamma_{\chi_{c1}} $)[MeV] & $g_{D\bar{D}^*}^{\chi_{c1}}$[GeV$^{-1/2}$] &$\vert \tilde{X}_{\chi_{c1}} \vert $  & $\tilde{Z}_{\chi_{c1}}$  \\ [2pt] \hline \hline
0.00   & -0.789 & 1.00  & 1.00& (3937.0, 0.0) &   0.00 &0.00  &1.00  \\ [2pt] 
0.05 & -0.781 & 0.90 &0.99  & (3937.42, 1.36)   & 0.03+0.11i   &   0.01  &  1.00+0.01i   \\[2pt]
0.10   & -0.758 & 0.89  & 0.98  & (3938.6, 5.54)    &0.07+0.21i  & 0.02 &0.99+0.02i     \\[2pt]
0.20  & -0.668 & 0.86  &0.92  & (3943.2, 23.9)   & 0.16+0.40i  & 0.09    &  0.96+0.09   \\[2pt]
0.30 & -0.517 &  0.82 &0.83  & (3949.1, 61.5)   &  0.28+0.56i   &  0.22   &  0.91+0.20i   \\[2pt]
0.40 & -0.305 & 0.77  &0.73  &(3951.7,  134.3)   & 0.44+0.68i    &  0.42   &  0.85+0.39i   \\[2pt]
0.50 & -0.033& 0.72 & 0.63 & (3930.6, 311.5)  & 0.70+0.82i    &0.94  &  0.67+0.88i   \\[2pt]
$\emph{d}^{\emph{crit}}$  & 0.000  & 0.71 &0.62 &(3923.5, 352.8)   & 0.74+0.86i    &   $>$1  &  0.59+1.01i   \\[2pt]
0.53 & 0.060  &  0.70 &  0.61  &(3081.0, 0.0)  &  0.47  &  $\vert \tilde{X}_{\chi_{c1}} \vert <1 $&  1.39   \\[2pt]
0.54 & 0.092  &  0.70 &  0.60  &(3457.4, 0.0)  & 0.94  &  $\vert \tilde{X}_{\chi_{c1}} \vert <1 $& 2.24  \\[2pt]
0.60 & 0.299  & 0.67 & 0.55 & (3795.6, 0.0)  & 0.87    & $\vert \tilde{X}_{\chi_{c1}} \vert <1$ & 1.86  \\[2pt]
1.00 &  2.233   & 0.49  &0.30  & (3862.4, 0.0) & 0.40   &  $\vert \tilde{X}_{\chi_{c1}} \vert <1 $ & 1.24     \\[2pt]
2.00 & 11.297  & 0.28  &  0.10  & (3870.0, 0.0) & 0.19    &  $\vert \tilde{X}_{\chi_{c1}} \vert <1 $&  1.07  \\[2pt]
4.00 & 47.554  & 0.15  &  0.03  & (3871.3, 0.0) & 0.09    &  $\vert \tilde{X}_{chi_{c1}} \vert <1 $ &  1.02  \\[2pt] \hline \hline
\end{tabular}    }}
\caption{ \small{ For $m^0_{c\bar{c}} $= 3937 MeV bare charmonium mass, dressed mass value of the $\chi_{1c}(2P)$ and its other properties as a function of $d$ ($ d^{crit}=0.510903$ fm$^{1/2}$).}} \label{tab:6}
\end{table}

\begin{table}
\centering{
\scalebox{0.8}{ 
\begin{tabular}{c  |c  c  c  | c  c  c c}
d [fm$^{1/2}$] & C$_{0X}$ [$fm^{2}$] & $g_{D\bar{D}^*}^{X(3872)}$[GeV$^{-1/2}$] & $\tilde{X}_{X(3872)}$ & ($m_{\chi_{c1}},\Gamma_{\chi_{c1}} $)[MeV] & $g_{D\bar{D}^*}^{\chi_{c1}}$[GeV$^{-1/2}$] &$\vert \tilde{X}_{\chi_{c1}} \vert $  & $\tilde{Z}_{\chi_{c1}}$  \\ [2pt] \hline \hline
0.00 & -0.789 & 1.00  &1.00  & (3947.0, 0.0) &  0.00 &0. 00 &1.00  \\ [2pt] 
0.05 & -0.782 & 0.90 & 1.00 & (3947.4, 1.3)   & 0.04 + 0.10i & 0.01 & 0.99 + 0.01i \\[2pt]
0.10 & -0.762 &0.89  & 0.98  & (3948.6, 5.1)  &0.08 + 0.19i &0.02 &  0.99 + 0.02i \\[2pt]
0.20  & -0.683 &0.87  & 0.93 & (3953.0, 21.8) &  0.16 + 0.37i & 0.08 &  0.97 + 0.07i\\ [2pt]
0.30 & -0.553 &  0.84 & 0.86 & (3959.3, 54.7) & 0.27 + 0.52i &  0.18 &  0.93 + 0.17i\\ [2pt]
0.40 & -0.369 & 0.80  &0.78  &(3965.2,  115.0)  & 0.42 + 0.63i & 0.34 &0.88 + 0.31i  \\ [2pt]
0.50 & -0.134  & 0.75 &0.70 &(3963.3, 237.0)  & 0.61 + 0.73i &  0.64 & 0.78 + 0.60i  \\ [2pt] 
$\emph{d}^{\emph{crit}}$ & 0.000  & 0.73 &0.66 &(3953.8, 365.3)  & 0.74 + 0.83i & 0.99 & 0.61 + 0.91i \\ [2pt]
0.60 & 0.155  & 0.71 & 0.61 & (3628.0, 0.0)& 1.00  & $\vert \tilde{X}_{\chi_{c1}} \vert <1$& 2.24 \\ [2pt]  
1.00 & 1.832 & 0.54  &0.37  & (3858.0, 0.0) & 0.46 &$\vert \tilde{X}_{\chi_{c1}} \vert <1 $& 1.30 \\ [2pt]  
2.00 & 9.692  & 0.32  & 0.13  & (3869.4, 0.0)& 0.22  & $\vert \tilde{X}_{\chi_{c1}} \vert <1 $& 1.09  \\ [2pt] 
4.00 & 41.135  & 0.17  & 0.03  & (3871.2, 0.0)& 0.11  &$\vert \tilde{X}_{\chi_{c1}} \vert <1 $& 1.03  \\ [2pt] 
6.00 & 93.538  & 0.11  & 0.02  & (3871.5, 0.0)& 0.07  & $\vert \tilde{X}_{\chi_{c1}} \vert <1 $& 1.01  \\ [2pt] 
 \hline \hline
\end{tabular}    }}
\caption{ \small{  For $m^0_{c\bar{c}} = $ 3947 MeV bare charmonium mass, dressed mass value of the $\chi_{1c}(2P)$ and its other properties as a function of $d$ ($ d^{crit}=0.548624$ fm$^{1/2}$). }} \label{tab:7}
\end{table}

\begin{table}
\centering{
\scalebox{0.8}{ 
\begin{tabular}{c  |c  c  c  | c  c  c c}
d [fm$^{1/2}$] & C$_{0X}$ [$fm^{2}$] & $g_{D\bar{D}^*}^{X(3872)}$[GeV$^{-1/2}$] & $\tilde{X}_{X(3872)}$ & ($m_{\chi_{c1}},\Gamma_{\chi_{c1}} $)[MeV] & $g_{D\bar{D}^*}^{\chi_{c1}}$[GeV$^{-1/2}$] &$\vert \tilde{X}_{\chi_{c1}} \vert $  & $\tilde{Z}_{\chi_{c1}}$  \\ [2pt] \hline \hline
0.00   & -0.789 & 1.00  &1.00 & (3953.0, 0.0)   &  0.00  & 0.00 & 1.00 \\ [2pt] 
0.05 & -0.783 & 0.90 & 1.00 & (3953.4, 1.2)    & 0.04+0.09i   &0.01  & 1.00+0.00i \\[2pt]
0.10 & -0.764& 0.89  &  0.99  & (3954.5, 4.9)    &  0.08+0.18i  &0.02  & 0.99+0.02i   \\[2pt]
0.20  & -0.692 & 0.88  & 0.94& (3958.8, 20.7)   & 0.17+0.35i  & 0.07   &0.97+0.07i   \\ [2pt]
0.30 & -0.570 &  0.85 &0.88  & (3965.3, 51.4)  & 0.27+0.50i  & 0.16   & 0.94+0.15i \\ [2pt]
0.40 & -0.400 & 0.81  & 0.81  &(3972.2,  106.2)   & 0.41+0.61i  &0.30    & 0.89+0.28i \\ [2pt]
0.50 & -0.182  & 0.77 &0.73 &(3975.2, 210.5)  & 0.58+0.70i  & 0.54   & 0.81+0.51i \\ [2pt]    
$\emph{d}^{\emph{crit}}$ & 0.000  & 0.74  &0.67 &(3970.6, 370.1)   & 0.74+0.82i  &0.95    &0.61+0.87i \\ [2pt]
0.60 & 0.085  & 0.73 &0.65  & (3185.6, 0.0)  & 0.50 &  $\vert \tilde{X}_{\chi_{c1}} \vert <1 $& 1.40 \\ [2pt]  
0.65 & 0.237  & 0.70 &0.61  & (3716.5, 0.0)  & 0.93 & $\vert \tilde{X}_{\chi_{c1}} \vert <1 $& 2.00 \\ [2pt]   
1.00 & 1.638  & 0.57 &0.40  & (3854.6, 0.0)  & 0.50 & $\vert \tilde{X}_{\chi_{c1}} \vert <1 $& 1.33 \\ [2pt]   
2.00 & 8.919  & 0.34 &0.14 & (3868.9, 0.0)  & 0.24 & $\vert \tilde{X}_{\chi_{c1}} \vert <1 $& 1.10 \\ [2pt]  
4.00 & 38.041  & 0.18 &0.04  & (3871.1, 0.0)  & 0.12 & $\vert \tilde{X}_{\chi_{c1}} \vert <1 $& 1.03 \\ [2pt] 
6.00 & 86.578 & 0.12 &0.02  & (3871.4, 0.0)  & 0.08 & $\vert \tilde{X}_{\chi_{c1}} \vert <1 $& 1.01 \\ [2pt]   \hline\hline
\end{tabular}    }}
\caption{ \small{  For $m^0_{c\bar{c}} =$ 3953 MeV bare charmonium mass, dressed mass value of the $\chi_{1c}(2P)$ and its other properties as a function of $d$ ($ d^{crit}=$0.57006 fm$^{1/2}$). }} \label{tab:8} 
\end{table}

\clearpage

Numerical results for an UV cutoff $\Lambda=0.5$ GeV:

\begin{table}[h]
\centering{
\scalebox{0.8}{ 
\begin{tabular}{c  |c  c  c  | c  c  c c}
d [fm$^{1/2}$] & C$_{0X}$ [$fm^{2}$] & $g_{D\bar{D}^*}^{X(3872)}$[GeV$^{-1/2}$] & $\tilde{X}_{X(3872)}$ & ($m_{\chi_{c1}},\Gamma_{\chi_{c1}} $)[MeV] & $g_{D\bar{D}^*}^{\chi_{c1}}$[GeV$^{-1/2}$] &$\vert \tilde{X}_{\chi_{c1}} \vert $  & $\tilde{Z}_{\chi_{c1}}$  \\ [2pt] \hline \hline
0.01  & -1.941 & 1.052 &0.995 & (3865.0, 0.21) &  -0.02i &0.001  &1.00  \\ [2pt] 
0.05 & -2.011& 0.997 & 0.894 & (3865.3, 0.11)  &   0.04-0.02i   &  0.013&  1.01-0.01i \\[2pt]
0.10  & -2.232 & 0.868  &0.677  & (3866.1, 1.65)  &  0.05-0.20i  & 0.048 & 1.05-0.01i  \\[2pt]
0.20 & -3.118& 0.619& 0.344 & (3868.5, 2.80)   & 0.17-0.25i   & 0.110   & 1.11+0.03i   \\[2pt]
0.40 & -6.657 & 0.359  & 0.116 &(3870.7, 1.34) & 0.18-0.12i  & 0.070 & 1.04+0.06i\\[2pt]
1.00 & -31.434 & 0.151  &  0.021 & (3871.5, 0.25)& 0.08-0.04i  &  0.015 & 1.01+0.01i  \\[2pt]
3.00 & -267.4& 0.051 & 0.002   & (3871.67, 0.03) & 0.03-0.01i   & 0.002 &   1.00 \\[2pt]

10.00 & -2951.5& 0.015  & 0.000   & (3871.69, 0.00) & 0.01  & 0.000  &   1.00 \\[2pt]

\hline \hline
\end{tabular} } }
\caption{\small{For $m^0_{c\bar{c}} = $ 3865 MeV bare charmonium mass, dressed mass value of the $\chi_{1c}(2P)$ and its other properties as a function of $d$. }} 
\label{tab:05}
\end{table}


\begin{table}[h]
\centering{
\scalebox{0.8}{ 
\begin{tabular}{c  |c  c  c  | c  c  c c}
d [fm$^{1/2}$] & C$_{0X}$ [$fm^{2}$] & $g_{D\bar{D}^*}^{X(3872)}$[GeV$^{-1/2}$] & $\tilde{X}_{X(3872)}$ & ($m_{\chi_{c1}},\Gamma_{\chi_{c1}} $)[MeV] & $g_{D\bar{D}^*}^{\chi_{c1}}$[GeV$^{-1/2}$] &$\vert \tilde{X}_{\chi_{c1}} \vert $  & $\tilde{Z}_{\chi_{c1}}$  \\ [2pt] \hline \hline
0.00  & -1.938 & 1.00 &1.00 & (3896.0, 0.0) &   0.00 &0.00  &1.00  \\ [2pt] 
0.05 & -1.917 & 1.05 & 0.99 & (3896.2, 0.4)   &  0.02-0.18i   &  0.006 &  0.99+0.04i     \\[2pt]
0.10  & -1.856 & 1.03  &0.96  & (3896.8, 1.8)    &  0.04-0.16i   & 0.025  & 0.98+0.02i    \\[2pt]
0.20 & -1.613& 0.98 & 0.87 & (3899.2, 7.95)   & 0.10+0.30i    & 0.09   & 0.93+0.07i    \\[2pt]
0.30 & -1.207 & 0.91 & 0.75 & (3903.1, 21.1)   & 0.19+0.4i    &  0.21  &  0.86+0.16i   \\[2pt]
0.40 & -0.639 & 0.83  & 0.63 &(3908.2, 48.3) & 0.33+0.47i  &  0.40& 0.78+0.33i \\[2pt]
$\emph{d}^{\emph{crit}}$ & 0.000 & 0.77 &0.53 & (3917.1, 105.9) &  0.48+0.55i   &  0.80 & 0.53+0.65i     \\[2pt]
0.50 & 0.091  & 0.76 & 0.52 &(3920.2, 118.4) & 0.49+0.57i  & 0.89  & 0.43+0.69i  \\[2pt]
0.70 & 2.039  & 0.63 & 0.36 &(3856.6, 0.0) & 0.35  & $\vert \tilde{X}_{\chi_{c1}} \vert <1 $  & 1.28 \\[2pt]
1.00 & 6.179  & 0.49  &  0.21& (3866.9, 0.0)&0.23   &$\vert \tilde{X}_{\chi_{c1}} \vert <1 $ & 1.13  \\[2pt]
2.00 & 30.53 & 0.27  & 0.06   & (3870.7, 0.0) & 0.113   & $\vert \tilde{X}_{\chi_{c1}} \vert <1 $   &   1.03 \\[2pt]
\hline \hline
\end{tabular} } }
\caption{\small{For $m^0_{c\bar{c}} = $ 3896 MeV bare charmonium mass, dressed mass value of the $\chi_{1c}(2P)$ and its other properties as a function of $d$ ($ d^{crit}=0.488589 $ fm$^{1/2}$).}} 
\label{tab:9}
\end{table}
\begin{table}[h]
\centering{
\scalebox{0.8}{ 
\begin{tabular}{c  |c  c  c  | c  c  c c}
d [fm$^{1/2}$] & C$_{0X}$ [$fm^{2}$] & $g_{D\bar{D}^*}^{X(3872)}$[GeV$^{-1/2}$] & $\tilde{X}_{X(3872)}$ & ($m_{\chi_{c1}},\Gamma_{\chi_{c1}} $)[MeV] & $g_{D\bar{D}^*}^{\chi_{c1}}$[GeV$^{-1/2}$] &$\vert \tilde{X}_{\chi_{c1}} \vert $  & $\tilde{Z}_{\chi_{c1}}$  \\ [2pt] \hline \hline
0.00  & -1.938 & 1.00 &1.00 & (3910.0, 0.0) &   0.00 &0.00  &1.00  \\ [2pt] 
0.05 & -1.925& 1.05 & 0.99 & (3910.1, 0.3)   &  0.03-0.05i   &  0.003 &  0.99+0.02i     \\[2pt]
0.10  & -1.886 & 1.04  &0.98  & (3910.6, 1.4)    &  0.06-0.11i   & 0.013  & 0.99+0.01i    \\[2pt]
0.20 & -1.731& 1.024 & 0.94 & (3912.6, 5.8)   & 0.13+0.22i    & 0.05   & 0.97+0.04i    \\[2pt]
0.30 & -1.474 & 0.99 & 0.88 & (3916.0, 14.2)   & 0.21+0.30i    &  0.11  &  0.93+0.10i   \\[2pt]
0.40 & -1.113 & 0.94  & 0.81 &(3921.0, 28.4) & 0.29+0.37i  &  0.21& 0.88+0.18i \\[2pt]
0.50 & -0.650  & 0.90 & 0.73 &(3928.5, 52.3) & 0.39+0.42i  & 0.36  & 0.81+0.31i  \\[2pt]
$\emph{d}^{\emph{crit}}$ & 0.000 & 0.85 &0.64 & (3946.1, 100.5) &  0.51+0.48i   &  0.67 & 0.57+0.52i     \\[2pt]
0.70 & 0.586 & 0.80 & 0.58 &(3979.1, 139.1) & 0.52+0.54i  & 0.89  & 0.23+0.46i \\[2pt]
1.00 & 3.213  & 0.67  &  0.40& (3855.4, 0.0)&0.32   &$\vert \tilde{X}_{\chi_{c1}} \vert <1 $ & 1.24  \\[2pt]
2.00 & 18.665& 0.40  & 0.14   & (3869.2, 0.0) & 0.174   & $\vert \tilde{X}_{\chi_{c1}} \vert <1 $   &   1.07 \\[2pt]
3.00 & 44.419& 0.28  & 0.07   & (3870.6, 0.0) & 0.117   & $\vert \tilde{X}_{\chi_{c1}} \vert <1 $   &   1.03 \\[2pt]
\hline \hline
\end{tabular} } }
\caption{\small{For $m^0_{c\bar{c}} = $ 3910 MeV bare charmonium mass, dressed mass value of the $\chi_{1c}(2P)$ and its other properties as a function of $d$ ($ d^{crit}=0.613349 $ fm$^{1/2}$).  }} 
\label{tab:10}
\end{table}
\begin{table}[h]
\centering{
\scalebox{0.8}{ 
\begin{tabular}{c  |c  c  c  | c  c  c c}
d [fm$^{1/2}$] & C$_{0X}$ [$fm^{2}$] & $g_{D\bar{D}^*}^{X(3872)}$[GeV$^{-1/2}$] & $\tilde{X}_{X(3872)}$ & ($m_{\chi_{c1}},\Gamma_{\chi_{c1}} $)[MeV] & $g_{D\bar{D}^*}^{\chi_{c1}}$[GeV$^{-1/2}$] &$\vert \tilde{X}_{\chi_{c1}} \vert $  & $\tilde{Z}_{\chi_{c1}}$  \\ [2pt] \hline \hline
0.00  & -1.938 & 1.00 &1.00 & (3925.0, 0.0) &   0.00 &0.00  &1.00  \\ [2pt] 
0.05 & -1.928& 1.05 & 0.99 & (3925.1, 0.2)   &  0.03-0.04i   &  0.002 &  0.99+0.02i     \\[2pt]
0.10  & -1.900 & 1.05  &0.99  & (3925.5, 1.1)    &  0.07-0.08i   & 0.008  & 0.99+0.07i    \\[2pt]
0.20 & -1.789& 1.04 & 0.97 & (3927.3, 4.5)   & 0.13+0.16i    & 0.03   & 0.98+0.03i    \\[2pt]
0.30 & -1.604 & 1.02 & 0.93 & (3930.2, 10.7)   & 0.21+0.23i    &  0.07  &  0.96+0.06i   \\[2pt]
0.40 & -1.345 & 0.99  & 0.89 &(3934.7, 20.4) & 0.28+0.28i  &  0.14& 0.93+0.12i \\[2pt]
0.50 & -1.012  & 0.96 & 0.84 &(3941.0, 34.9) & 0.36+0.33i  & 0.23  & 0.88+0.20i  \\[2pt]
$\emph{d}^{\emph{crit}}$ & 0.000 & 0.89 &0.72 & (3970.9, 91.9) &  0.53+0.42i   &  0.59 & 0.61+0.44i     \\[2pt]
1.00 & 1.763  & 0.79  &  0.57& (3822.8, 0.0)&0.33   & $\vert \tilde{X}_{\chi_{c1}} \vert <1 $ & 1.26  \\[2pt]
2.00 & 12.868& 0.52  & 0.25   & (3866.4, 0.0) & 0.23   & $\vert \tilde{X}_{\chi_{c1}} \vert <1 $   &   1.12 \\[2pt]
4.00 & 57.286& 0.29  & 0.07   & (3870.6, 0.0) & 0.12   & $\vert \tilde{X}_{\chi_{c1}} \vert <1 $   &   1.04 \\[2pt]
\hline \hline
\end{tabular} } }
\caption{\small{For $m^0_{c\bar{c}} = $ 3925 MeV bare charmonium mass, dressed mass value of the $\chi_{1c}(2P)$ and its other properties as a function of $d$ ($ d^{crit}=0.723529 $ fm$^{1/2}$). }} 
\label{tab:11}
\end{table}
\begin{table}[h]
\centering{
\scalebox{0.8}{ 
\begin{tabular}{c  |c  c  c  | c  c  c c}
d [fm$^{1/2}$] & C$_{0X}$ [$fm^{2}$] & $g_{D\bar{D}^*}^{X(3872)}$[GeV$^{-1/2}$] & $\tilde{X}_{X(3872)}$ & ($m_{\chi_{c1}},\Gamma_{\chi_{c1}} $)[MeV] & $g_{D\bar{D}^*}^{\chi_{c1}}$[GeV$^{-1/2}$] &$\vert \tilde{X}_{\chi_{c1}} \vert $  & $\tilde{Z}_{\chi_{c1}}$  \\ [2pt] \hline \hline
0.00  & -1.938 & 1.00 &1.00 & (3937.0, 0.0) &   0.00 &0.00  &1.00  \\ [2pt] 
0.05 & -1.930& 1.05 & 0.99 & (3937.1, 0.2)   &  0.03-0.03i   &  0.001 &  0.99+0.00i     \\[2pt]
0.10  & -1.907 & 1.05  &0.99  & (3937.5, 0.9)    &  0.07-0.06i   & 0.006  & 0.99+0.00i    \\[2pt]
0.20 & -1.817& 1.04 & 0.98 & (3939.1, 3.8)   & 0.13+0.12i    & 0.02   & 0.98+0.02i    \\[2pt]
0.30 & -1.665 & 1.03 & 0.95 & (3941.8, 9.0)   & 0.20+0.18i    &  0.06  &  0.97+0.05i   \\[2pt]
0.40 & -1.454 & 1.01  & 0.92 &(3945.9, 16.7) & 0.27+0.23i  &  0.11& 0.94+0.01i \\[2pt]
0.50 & -1.182 & 0.99 & 0.89 &(3951.5, 27.7) & 0.34+0.27i  & 0.17  & 0.91+0.15i  \\[2pt]
$\emph{d}^{\emph{crit}}$ & 0.000 & 0.91 &0.75 & (3988.2, 84.8) &  0.53+0.37i   &  0.54 & 0.62+0.39i     \\[2pt]
1.00 & 1.083  & 0.86  &  0.66& (3762.3, 0.0)&0.23   & $\vert \tilde{X}_{\chi_{c1}} \vert <1 $ & 1.16  \\[2pt]
1.50 & 4.860& 0.72  & 0.47   & (3852.0, 0.0) & 0.31   & $\vert \tilde{X}_{\chi_{c1}} \vert <1 $   &   1.21 \\[2pt]
2.00 & 10.1478& 0.61  & 0.33   & (3869.0, 0.0) & 0.33   & $\vert \tilde{X}_{\chi_{c1}} \vert <1 $   &   1.48 \\[2pt]
4.00 & 46.404& 0.35  & 0.11   & (3869.9, 0.0) & 0.14   & $\vert \tilde{X}_{\chi_{c1}} \vert <1 $   &   1.05 \\[2pt]
\hline \hline
\end{tabular} } }
\caption{\small{For $m^0_{c\bar{c}} = $ 3937 MeV bare charmonium mass, dressed mass value of the $\chi_{1c}(2P)$ and its other properties as a function of $d$ ($ d^{crit}=0.800832 $ fm$^{1/2}$).  }} 
\label{tab:12}
\end{table}
\begin{table}[h]
\centering{
\scalebox{0.8}{ 
\begin{tabular}{c  |c  c  c  | c  c  c c}
d [fm$^{1/2}$] & C$_{0X}$ [$fm^{2}$] & $g_{D\bar{D}^*}^{X(3872)}$[GeV$^{-1/2}$] & $\tilde{X}_{X(3872)}$ & ($m_{\chi_{c1}},\Gamma_{\chi_{c1}} $)[MeV] & $g_{D\bar{D}^*}^{\chi_{c1}}$[GeV$^{-1/2}$] &$\vert \tilde{X}_{\chi_{c1}} \vert $  & $\tilde{Z}_{\chi_{c1}}$  \\ [2pt] \hline \hline
0.00  & -1.938 & 1.00 &1.00 & (3947.0, 0.0) &   0.00 &0.00  &1.00  \\ [2pt] 
0.05 & -1.931& 1.05 & 0.99 & (3947.1, 0.2)   &  0.03-0.02i   &  0.001 &  0.99+0.00i     \\[2pt]
0.10  & -1.911 & 1.05  &0.99  & (3947.5, 0.8)    &  0.06-0.05i   & 0.005  & 0.99+0.00i    \\[2pt]
0.20 & -1.833& 1.04 & 0.98 & (3949.0, 3.4)   & 0.13+0.10i    & 0.02   & 0.98+0.02i    \\[2pt]
0.30 & -1.702 & 1.03 & 0.96 & (3951.8, 7.9)   & 0.20+0.15i    &  0.05  &  0.97+0.04i   \\[2pt]
0.40 & -1.518 & 1.02  & 0.94 &(3955.3, 14.5) & 0.26+0.19i  &  0.10& 0.95+0.08i \\[2pt]
0.50 & -1.282 & 1.00 & 0.91 &(3960.6, 23.6) & 0.33+0.23i  & 0.14  & 0.93+0.12i  \\[2pt]
 0.70 & -0.653 & 0.968 & 0.84 &(3977.5, 50.9) & 0.45+0.28i  & 0.31  & 0.82+0.25i  \\[2pt]
$\emph{d}^{\emph{crit}}$ & 0.000 & 0.93 &0.78 & (4001.6, 79.0) &  0.53+0.33i   &  0.50 & 0.64+0.36i     \\[2pt]
1.00 & 0.682  & 0.90  &  0.72& (3660.8, 0.0)&0.12   & $\vert \tilde{X}_{\chi_{c1}} \vert <1 $ & 1.05  \\[2pt]
2.00 & 8.543& 0.66  & 0.39   & (3859.3, 0.0) & 0.28   & $\vert \tilde{X}_{\chi_{c1}} \vert <1 $   &   1.18 \\[2pt]
4.00 & 41.13& 0.16  & 0.03   & (3871.1, 0.0) & 0.11   & $\vert \tilde{X}_{\chi_{c1}} \vert <1 $   &   1.02 \\[2pt]
\hline \hline
\end{tabular} } }
\caption{\small{For $m^0_{c\bar{c}} = $ 3947 MeV bare charmonium mass, dressed mass value of the $\chi_{1c}(2P)$ and its other properties as a function of $d$ ($ d^{crit}=0.859959 $ fm$^{1/2}$). }} 
\label{tab:13}
\end{table}
\begin{table}[h]
\centering{
\scalebox{0.8}{ 
\begin{tabular}{c  |c  c  c  | c  c  c c}
d [fm$^{1/2}$] & C$_{0X}$ [$fm^{2}$] & $g_{D\bar{D}^*}^{X(3872)}$[GeV$^{-1/2}$] & $\tilde{X}_{X(3872)}$ & ($m_{\chi_{c1}},\Gamma_{\chi_{c1}} $)[MeV] & $g_{D\bar{D}^*}^{\chi_{c1}}$[GeV$^{-1/2}$] &$\vert \tilde{X}_{\chi_{c1}} \vert $  & $\tilde{Z}_{\chi_{c1}}$  \\ [2pt] \hline \hline
0.00  & -1.938 & 1.00 &1.00 & (3953.0, 0.0) &   0.00 &0.00  &1.00  \\ [2pt] 
0.05 & -1.931& 1.05 & 0.99 & (3953.1, 0.2)   &  0.03-0.02i   &  0.001 &  0.99+0.00i     \\[2pt]
0.10  & -1.913 & 1.05  &0.99  & (3953.5, 0.8)    &  0.06-0.04i   & 0.005  & 0.99+0.00i    \\[2pt]
0.20 & -1.840& 1.04 & 0.98 & (3954.9, 3.2)   & 0.13+0.10i    & 0.02   & 0.99+0.01i    \\[2pt]
0.30 & -1.719 & 1.04 & 0.97 & (3957.4, 7.3)   & 0.19+0.13i    &  0.04  &  0.98+0.04i   \\[2pt]
0.40 & -1.549 & 1.03  & 0.95 &(3961.1, 13.4) & 0.26+0.17i  &  0.08& 0.96+0.07i \\[2pt]
0.50 & -1.331 & 1.01 & 0.92 &(3966.1, 21.7) & 0.32+0.20i  & 0.13  & 0.93+0.11i  \\[2pt]
 0.70 & -0.748 & 0.979 & 0.86 &(3981.9, 45.8) & 0.44+0.26i  & 0.28  & 0.84+0.23i  \\[2pt]
$\emph{d}^{\emph{crit}}$ & 0.000 & 0.94 &0.79 & (4009.2, 75.7) &  0.53+0.31i   &  0.48 & 0.65+0.34i     \\[2pt]
1.00 & 0.489  & 0.91  &  0.75 & (3550.4, 0.0)&0.06   & $\vert \tilde{X}_{\chi_{c1}} \vert <1 $ & 1.01  \\[2pt]
2.00 & 7.769& 0.69  & 0.43   & (3856.6, 0.0) & 0.29   & $\vert \tilde{X}_{\chi_{c1}} \vert <1 $   &   1.19 \\[2pt]
4.00 & 36.89& 0.42  & 0.16   & (3868.9, 0.0) & 0.17   & $\vert \tilde{X}_{\chi_{c1}} \vert <1 $   &   1.08 \\[2pt]
\hline \hline
\end{tabular} } }
\caption{\small{For $m^0_{c\bar{c}} = $ 3953 MeV bare charmonium mass, dressed mass value of the $\chi_{1c}(2P)$ and its other properties as a function of $d$ ($ d^{crit}=0.893559 $ fm$^{1/2}$). }} 
\label{tab:14}
\end{table}

\clearpage

\bibliography{paper}

\end{document}